\documentclass[onecolumn]{IEEEtran}
\usepackage[latin9]{inputenc}
\usepackage{float}
\usepackage{dsfont}
\usepackage{amsmath}
\usepackage{amsthm}
\usepackage{amssymb}
\usepackage{graphicx}
\PassOptionsToPackage{normalem}{ulem}
\usepackage{ulem}
\usepackage[unicode=true,pdfusetitle,
 bookmarks=true,bookmarksnumbered=false,bookmarksopen=false,
 breaklinks=false,pdfborder={0 0 0},pdfborderstyle={},backref=false,colorlinks=false]
 {hyperref}

\makeatletter

\floatstyle{ruled}
\newfloat{algorithm}{tbp}{loa}
\providecommand{\algorithmname}{Algorithm}
\floatname{algorithm}{\protect\algorithmname}

\theoremstyle{plain}
\newtheorem{notation}{\protect\notationname}
\theoremstyle{definition}
\newtheorem{defn}{\protect\definitionname}
\theoremstyle{definition}
\newtheorem{problem}{\protect\problemname}
\theoremstyle{remark}
\newtheorem{rem}{\protect\remarkname}
\theoremstyle{plain}
\newtheorem{thm}{\protect\theoremname}
\theoremstyle{plain}
\newtheorem{lem}{\protect\lemmaname}
\theoremstyle{plain}
\newtheorem{cor}{\protect\corollaryname}
\theoremstyle{plain}
\newtheorem{prop}{\protect\propositionname}

\@ifundefined{date}{}{\date{}}
\makeatother

\providecommand{\corollaryname}{Corollary}
\providecommand{\definitionname}{Definition}
\providecommand{\lemmaname}{Lemma}
\providecommand{\notationname}{Notation}
\providecommand{\problemname}{Problem}
\providecommand{\propositionname}{Proposition}
\providecommand{\remarkname}{Remark}
\providecommand{\theoremname}{Theorem}

\begin{document}
\title{Statistical Approach to Detection of Attacks for Stochastic Cyber-Physical
Systems}
\author{Damián Marelli$^{1}$, Tianju Sui$^{2,\dagger}$, Minyue Fu$^{3}$,
\emph{Fellow IEEE} and Renquan Lu$^{4}$\thanks{$^{1}$Damián Marelli is with the School of Automation, Guangdong
University of Technology, Guangzhou, China, and with the French Argentine
International Center for Information and Systems Sciences, National
Scientific and Technical Research Council, Argentina. Email: \texttt{\footnotesize{}Damian.Marelli@newcastle.edu.au}{\footnotesize{}.}}\thanks{$^{2}$Tianju Sui is with the School of Control Science and Engineering,
Dalian University of Technology, Dalian, China. Email:{\footnotesize{}
}\texttt{\footnotesize{}suitj@mail.dlut.edu.cn.}}\thanks{$^{3}$Minyue Fu is with the School of Electrical Engineering and
Computer Science, University of Newcastle, Callaghan, NSW 2308, Australia.
Email:{\footnotesize{} }\texttt{\footnotesize{}Minyue.Fu@newcastle.edu.au.}}\thanks{$^{4}$Renquan Lu is with the School of Automation, and also with
the Guangdong Provincial Key Laboratory of Intelligent Decision and
Cooperative Control, Guandong University of Technology, Guangzhou,
China. Email:{\footnotesize{} }\texttt{\footnotesize{}rqliu@gdut.edu.cn.}}\thanks{$\dagger$Corresponding author.}\thanks{This work was supported by National Natural Science Foundation of
China (61803068), China Postdoctoral Science Foundation(2017M621134),
Australian Research Council under Grant DP200103507, and Argentinean
Agency for Scientific and Technological Promotion (PICT- 201-0985).}}
\maketitle
\begin{abstract}
We study the problem of detecting an attack on a stochastic cyber-physical
system. We aim to treat the problem in its most general form. We start
by introducing the notion of asymptotically detectable attacks, as
those attacks introducing changes to the system's output statistics
which persist asymptotically. We then provide a necessary and sufficient
condition for asymptotic detectability. This condition preserves generality
as it holds under no restrictive assumption on the system and attacking
scheme. To show the importance of this condition, we apply it to detect
certain attacking schemes which are undetectable using simple statistics.
Our necessary and sufficient condition naturally leads to an algorithm
which gives a confidence level for attack detection. We present simulation
results to illustrate the performance of this algorithm.
\end{abstract}

\section{Introduction\label{sec:intro}}

A Cyber-physical systems (CPS) is a physical system which is monitored
or controlled via a communication channel. It finds a wide range of
applications such as traffic signal systems~\cite{traffic2015},
health care~\cite{Sastry2008chanllenge}, energy manufacturing~\cite{chen2010stuxnet},
power system~\cite{DeMarco1996,Sridhar2012,Mohsenian2011,dan2010,Bullo2011},
the water industry~\cite{Amin2010,Eliades2010}. A CPS is prone to
attacks in the form of signals injected through the communication
link~\cite{Sastry2008chanllenge}. These attacks are known to have
caused a number of serious accidents around the world~\cite{Farwell2011,Richards2008,Conti2010blackout,slay2007water,Kuvshinkova2003SQL}.
They have generated urgency for detecting such attacks.

In principle, an attack may be regarded as a system fault. This permits
using methods for fault tolerant control, such as robust statistics~\cite{Huber2011},
robust control~\cite{Zhou1996} and failure detection and identification~\cite{Willsky1976}.
However, the essential difference between an attack and a fault is
that the design of the former aims at making it difficult for detection.
For example, Liu \emph{et al} studied how to inject a stealthy input
into the measurement without being detected by the classical failure
detector~\cite{Liu2011}. Hence, methods for CPS attack detection
need to take special care of this difference.

Early works on CPS attack detection rely on certain prior knowledge
of the attacker's model. Among these methods, we find: The work in~\cite{Amin2009},
deals with a kind of attack called denial of service. The works~\cite{Liu2011}
and~\cite{Mo2010false} concentrate on false data injection attacks
against state estimation. The authors of~\cite{Teixeira2010} introduced
stealthy deception attacks, which consist in manipulating the measurements
to be processed by a power system state estimator in such a manner
that the resulting systematic errors introduced by the adversary are
either undetected or only partially detected by a bad data detection
method. In~\cite{Mo2009replay}, the effect of replay attacks is
studied. Smith investigated the behavior of control systems under
covert attacks~\cite{Smith2011}, where a malicious agent can access
the signals and information within the control loop and use them to
disrupt or compromise the controlled plant.

It is often unrealistic to assume that the defender has some knowledge
of the attacker's model. To address this concern, recent works have
studied the CPS attack problem without an attacking model assumption.
In this line, Pasqualetti \emph{et al} studied the problem of detectability,
by describing the undetectable/unidentificable attack class, consisting
of attacks not detectable by any kind of detection method~\cite{Pasqualetti2013}.
Using this concept, they studied in~\cite{Pasqualetti2013,Pasqualetti2012part2}
the design of centralized and distributed attack detection methods.
However, this approach is limited to systems without process and measurement
noises. The study of systems involving random noises is much more
challenging, since these systems present more ambiguities where attacks
can be hidden.

Concerning the detection of attacks in systems with noise, Mo and
Sinopoli~\cite{mo2016performance} analyzed the estimation error
introduced by an attack which is not detected by a $\chi^{2}$ failure
detector. They also studied in~\cite{mo2015secure} the attacks on
scalar systems with multiple sensors. In~\cite{bai2017kalman}, the
authors introduce the notion of strictly stealthy and $\epsilon$-stealthy
attacks, and bound the performance deterioration achievable by such
attacks.

In this work we move a step forward in the research line described
above. As in~\cite{mo2016performance,mo2015secure,bai2017kalman},
we also study systems with noise. We start by introducing the notion
of asymptotically stealthy attacks, as an extension of the definition
of strict stealthiness given in~\cite{bai2017kalman}, to the case
where the system and attack are non-stationary. More precisely, strict
stealthiness means that the attack does not change the output statistics.
Therefore, it cannot be detected by any method using statistical knowledge
of the system's output. However, if the detector only knows a single
realization of the system's output, this definition is too restrictive.
We relax this condition, by defining an attack to be asymptotically
stealthy if the changes it induces on output statistics vanish asymptotically.
We then define an attack to be asymptotically detectable if it is
not asymptotically stealthy. Some rigorous statistical setup is required
to make our definition precise.

To make the notion of asymptotic detectability verifiable, we provide
a necessary and sufficient condition for it. This condition is expressed
in terms of certain statistical properties, which are in principle
testable using the knowledge of a single realization. The condition
is given without requiring any assumption on the attacking model,
and, under mild regularity conditions, is valid for the general case
in which the system being attacked is time-varying, non-linear and
with non-Gaussian noises. We also specialize this condition for the
case of stationary linear system with Gaussian noises.

To appreciate the importance of the introduced notion of asymptotic
detectability and the provided necessary and sufficient condition,
we give two examples of attacks, which cannot be detected by checking
commonly used statistics, but are instead detected by checking our
condition.

In view of our main result, testing that an attack is detectable requires
verifying a condition which is numerically intractable. To fill this
gap, we derive a practically feasible detection algorithm. While we
do so for the case in which the system is linear and Gaussian, the
algorithm can be readily extended to arbitrary non-linear non-Gaussian
systems. Also, while the class of attacks that can be detected by
this algorithm is smaller than the class of asymptotically detectable
attacks, the difference between these two classes can be made arbitrarily
small by sufficiently increasing the complexity of the algorithm.

The rest of this paper is organized as follows. In Section~\ref{sec:Preliminaries}
we introduce the required statistical background. In Section~\ref{sec:problem}
we describe the attack detection problem. In Section~\ref{sec:strong}
we introduce the definition of asymptotically detectable attacks.
In Section~\ref{subsec:condition} we give a necessary and sufficient
condition for asymptotic detectability. More precisely, in Section~\ref{subsec:general-case}
we treat the general case, and in Section~\ref{subsec:linear-Gaussian}
we specialize this result for the case of stationary linear system
with Gaussian noises. In Section~\ref{sec:relevant}, we discuss
attack examples which cannot be detected by checking other simpler
conditions. In Section~\ref{sec:algorithm} we use our condition
to derive the detection algorithm. In Section~\ref{sec:simulate}
we use simulations to illustrate the superiority of our algorithm
for detecting attacks that cannot be detected by other simpler methods.
Finally, concluding remarks are stated in Section~\ref{sec:conclusion}.

\section{Preliminaries\label{sec:Preliminaries}}
\begin{notation}
We use $\mathbb{N}$ to denote the set of natural numbers, $\mathbb{Z}$
to denote the set of integers, $\mathbb{R}_{+}$ to denote the set
of non-negative real numbers and $\bar{\mathbb{R}}=\mathbb{R}\cup\{-\infty,\infty\}$
to denote the set of extended real numbers. For a vector $x$ we use
$\left[x\right]_{i}$ to denote its $i$-th entry and for a matrix
$X$ we use $\left[X\right]_{i,j}$ to denote its $(i,j)$-th entry.
For a vector or matrix $X$, we use $X^{\top}$ to denote its transpose.
For vectors $x$ and $y$, $x\prec y$ ($x\preceq y$) means that
$\left[x\right]_{i}<\left[y\right]_{i}$ ($\left[x\right]_{i}\leq\left[y\right]_{i}$),
for all $i$, and $z=x\wedge y$ denotes the vector with entries $[z]_{i}=\min(x_{i},y_{i})$.
We use $\mathbf{1}$ to denote a column vector of ones, $\mathbf{I}$
to denote the identity matrix, $q$ to denote the forward-shift operator,
i.e., $(qx)_{t}=x_{t+1}$, and $\mathds{1}_{S}$ to denote the indicator
function of the set $s$, i.e., $\mathds{1}_{S}(x)=1$, if $x\in S$
and $0$ otherwise. We also use $\phi_{\mu,\Sigma}$ and $\Phi_{\mu,\Sigma}$
to denote the probability density function (PDF) and cumulative distribution
function (CDF), respectively, of the normal distribution with mean
$\mu$ and covariance matrix $\Sigma$.
\end{notation}
\begin{notation}
Let $\mathbb{S}^{D}=\left(\mathbb{R}^{D}\right)^{\mathbb{N}}$ denote
the set of all sequences $x=\left(x_{t}\in\mathbb{R}^{D}:t\in\mathbb{N}\right)$
indexed by the natural numbers. Let $\mathcal{\mathcal{S}}^{D}$ denote
the sigma algebra on $\mathbb{S}^{D}$ generated by the \emph{cylinder
sets} 
\begin{eqnarray*}
C_{t}\left(A\right) & = & \left\{ x\in\mathbb{S}^{D}:x_{t}\in A\right\} ,\quad\text{for all }t\in\mathbb{N},A\subseteq\mathbb{R}^{D}.
\end{eqnarray*}
\end{notation}
\begin{defn}
Let $\left(\Omega,\mathcal{A},\mathbb{P}\right)$ denote a probability
space. A $D$-dimensional random process is a map $x:\Omega\rightarrow\mathbb{S}^{D}$,
measurable with respect to $\mathcal{A}$ and $\mathcal{S}^{D}$.
Its probability distribution $p$ is 
\[
p\left(x\in S\right)=\mathbb{P}\circ x^{-1}\left(S\right)\text{ for all }S\in\mathcal{S}^{D}.
\]
A random process $x$ is said to be \emph{asymptotically mean stationary
(AMS)}~\cite[S 7.3]{Gray200908} if 
\[
\lim_{T\rightarrow\infty}\frac{1}{T}\sum_{t=1}^{T}p\left(q^{t}x\in S\right)\text{ exists for all }S\in\mathcal{S}^{D}.
\]
In this case, the associated stationary probability distribution $\bar{p}$
is defined by 
\begin{equation}
\bar{p}\left(x\in S\right)=\lim_{T\rightarrow\infty}\frac{1}{T}\sum_{t=1}^{T}p\left(q^{t}x\in S\right)\text{ for all }S\in\mathcal{S}^{D}.\label{eq:stat-dist}
\end{equation}
For a measurable map $f:\mathbb{S}^{D}\rightarrow\mathbb{R}$, we
define its stationary mean by
\[
\bar{\mathcal{E}}\left\{ f(x)\right\} =\int f(x)d\bar{p}.
\]
Also, $x$ is said to be \emph{ergodic}~\cite[S 7.7]{Gray200908}
if, for all $S\in\mathcal{S}^{D}$, 
\[
qS=S\Rightarrow p\left(qx\right)=0\text{ or }1.
\]
\end{defn}
AMS and ergodicity are properties which are stated in a rather technical
way. Roughly speaking, the AMS property is required for all limits
of sample averages to exist w.p.1. Also, ergodicity is required for
these limit values to be equal w.p.1. These statements are made precise
by the AMS ergodic theorem~\cite[Th. 8.1]{Gray200908}.

\section{Problem description\label{sec:problem}}

We have the following system in state-space form 
\begin{align}
x_{t+1} & =f_{t}\left(x_{t},w_{t}\right),\label{eq:ss1}\\
y_{t} & =g_{t}\left(x_{t},v_{t}\right),\label{eq:ss2}
\end{align}
where $f_{t}:\mathbb{R}^{D_{x}}\times\mathbb{R}^{D_{w}}\rightarrow\mathbb{R}^{D_{x}}$
and $g_{t}:\mathbb{R}^{D_{x}}\times\mathbb{R}^{D_{v}}\rightarrow\mathbb{R}^{D}$
are (measurable) non-linear time-varying functions, and the process
noise $w_{t}$, $t\in\mathbb{Z}$, and measurement noise $v_{t}$,
$t\in\mathbb{Z}$, are sequences of random vectors. We assume that
$y$ is AMS, ergodic, and its distribution $p\left(y\in S\right)$,
$S\in\mathcal{S}^{D}$ is absolutely continuous.

Consider an attacker, which interferes the measurement signal $y_{t}$,
replaces it with an attacking signal $z_{t}$, and sends $z_{t}$
instead of $y_{t}$ to the receiver. In order to treat the problem
in its full generality, we assume that $z_{t}$ is generated by an
arbitrary (possibly non-linear and non-stationary) measurable function
of the whole history of $y_{s}$ up to time $t$, i.e., 
\[
z_{t}=h_{t}\left(y_{s}:t\geq s\in\mathbb{N}\right).
\]

\begin{problem}
The attack detection problem consists in assessing whether or not
$z=y$. 
\end{problem}
\begin{defn}
\label{def:nominal} We say that the statistics of a random variable/process
are \emph{nominal} if they equal those which occur when $z=y$. The
probability law and expected value taken with respect to these statistics
are denoted by $p_{\star}\left(\cdot\right)$ and $\mathcal{E}_{\star}\left\{ \cdot\right\} $,
respectively. 
\end{defn}

\section{Asymptotic detectability\label{sec:strong}}

In order to assess whether $z=y$, all the information that we have
is a single realization of $z$ and the probability distribution of
$y$, i.e., $p(y\in S)$, for all $S\in\mathcal{S}^{D}$. The latter
equals the nominal probability distribution $p_{\star}(z\in S)$ of
$z$, i.e., when there is no attack. In~\cite{bai2017kalman}, under
the assumption that $y$ is stationary, an attack was defined to be
(strictly) stealthy if it satisfies a condition which is equivalent
to
\begin{equation}
p\left(z\in S\right)=p_{\star}\left(z\in S\right)\text{ for all }S\in\mathcal{S}.\label{eq:cond1}
\end{equation}
In the case in which $y$ is non-stationary, since we only know a
single realization of $z$, it is impossible to check~(\ref{eq:cond1}).
However, we can still hope to check whether the stationary distribution
of $z$ (cf.~(\ref{eq:stat-dist})) matches its nominal value, provided
both exist. This leads to our definition of asymptotic stealthiness
and asymptotic detectability listed below, which do not require the
existence of either stationary distribution.
\begin{defn}
\label{def:stochastic-detectability} We say that $z$ is asymptotically
stealthy if
\begin{equation}
\lim_{T\rightarrow\infty}\frac{1}{T}\sum_{t=1}^{T}\left[p\left(q^{t}z\in S\right)-p_{\star}\left(q^{t}z\in S\right)\right]=0\text{ for all }S\in\mathcal{S}.\label{eq:undetect-cond-0}
\end{equation}
Otherwise, we say that $z$ is asymptotically detectable.
\begin{rem}
An attack is asymptotically detectable if it causes a modification
in the probability distribution $p\left(q^{t}z\in S\right)$ which
is persistent over time. Hence, in particular, every finite-time attack
is asymptotically stealthy. Asymptotic detectability essentially means
that it is possible to detect the presence of the attack, with a confidence
that tends to one as the number of observed samples tends to infinity.
Obviously, in a practical setting, any method aiming at approximating
the Cesàro summation in~(\ref{eq:undetect-cond-0}) will be carried
out over a sliding time window of finite length. This permits detecting
finite-time attacks with a confidence that depends on the window length
and the duration of the attack. We derive one such methods in Section~\ref{sec:algorithm},
and discus the choice of the sliding window length in Remark~\ref{rem:time-horizon}.
\end{rem}
\end{defn}

\section{A necessary and sufficient condition for asymptotic detectability\label{subsec:condition}}

Our definition of asymptotic detectability is obviously impractical,
because it requires considering all possible test functions which
are integrable under the nominal distribution. In this section we
provide a necessary and sufficient condition for asymptotic detectability.
In Section~\ref{subsec:general-case} we do so for the general case
described in Section~\ref{sec:problem}, and in Section~\ref{subsec:linear-Gaussian}
we specialize this result for the case in which the system~(\ref{eq:ss1})-(\ref{eq:ss2})
is stationary, linear, and with Gaussian noises.

\subsection{The general case\label{subsec:general-case}}

Let $L\in\mathbb{N}$, $\zeta_{t}^{(L)}=\left[z_{t}^{\top},\cdots,z_{t+L-1}^{\top}\right]^{\top}$
and $\rho\in\mathbb{R}^{LD}$. Let also 
\begin{equation}
F_{\star}^{(L)}(\rho)=\bar{p}_{\star}\left(\zeta_{1}^{(L)}\preceq\rho\right),\label{eq:CDF}
\end{equation}
be the nominal CDF of $\zeta_{1}^{(L)}$. For each $T\in\mathbb{N}$,
let 
\begin{equation}
F_{T}^{(L)}\left(\rho\right)=\frac{1}{T}\sum_{t=1}^{T}\mathds{1}_{\left\{ \zeta_{t}^{(L)}\preceq\rho\right\} }(\rho),\label{eq:CDFapprox}
\end{equation}
be a sample approximation of the true stationary CDF $F^{(L)}(\rho)=\bar{p}\left(\zeta_{1}^{(L)}\preceq\rho\right)$.
The next theorem uses these definitions to provide a necessary and
sufficient condition for asymptotic stealthiness.
\begin{thm}
\label{thm:main} If $y$ is AMS, ergodic and its distribution is
absolutely continuous, the process $z$ is asymptotically stealthy
if and only if, for all $L\in\mathbb{N}$ and $\rho\in\mathbb{R}^{LD}$,
\begin{equation}
\lim_{T\rightarrow\infty}F_{T}^{(L)}\left(\rho\right)\overset{\text{w.p.1}}{=}F_{\star}^{(L)}(\rho).\label{eq:undetect-cond}
\end{equation}
\end{thm}
We devote the rest of the section to the proof of Theorem~\ref{thm:main}.
\begin{notation}
For $x,y\in\mathbb{S}^{D}$, we define the distance 
\[
d\left(x,y\right)=\sum_{t=1}^{\infty}2^{-t}\left(1\wedge\left|x_{t}-y_{t}\right|\right).
\]
As shown in~\cite[p. 241]{Billingsley199907}, under the topology
induced by $d$, $\mathbb{S}^{D}$ is separable and complete. For
a set $S\in\mathcal{S}^{D}$, we use $\bar{S}$ to denote its closure
under this topology. Also, we use $\partial S=\bar{S}\cap\overline{\left(\mathbb{S}^{D}\setminus S\right)}$
to denote the boundary of $S$.

For $L,T\in\mathbb{N}$, and $S\in\mathcal{S}^{D}$, let the measure
$\bar{p}_{T}\left(z\in\cdot\right):\mathcal{S}^{D}\rightarrow\mathbb{R}_{+}$
be defined by 
\[
\bar{p}_{T}\left(z\in S\right)=\frac{1}{T}\sum_{t=1}^{T}p\left(q^{t}z\in S\right).
\]

For $L\in\mathbb{N}$ and $\rho\in\bar{\mathbb{R}}^{LD}$, let $R_{\rho}^{L}\in\mathcal{\mathcal{S}}^{D}$
be the set 
\[
R_{\rho}^{L}=\left\{ x\in\mathbb{S}^{D}:\left[x_{1}^{\top},\cdots,x_{L}^{\top}\right]^{\top}\preceq\rho\right\} .
\]
\end{notation}
\begin{lem}
\label{lem:boundary} If $y$ is AMS and its distribution is absolutely
continuous, then $\bar{p}_{\star}\left(\check{z}\in\partial S\right)=0$,
for any $S\in\mathcal{S}^{D}$.
\end{lem}
\begin{IEEEproof}
We split the proof in steps:

Step 1) Let $\mathcal{R}=\left\{ R_{\rho}^{L}:L\in\mathbb{N},\rho\in\bar{\mathbb{R}}^{LD}\right\} $.
Let $L,M\in\mathbb{N}$, $\rho\in\bar{\mathbb{R}}^{LD}$ and $\lambda\in\bar{\mathbb{R}}^{MD}$.
Without loss of generality, suppose that $M\geq L$, and let $\tilde{\rho}=\left[\rho^{\top},\infty,\cdots,\infty\right]^{\top}$
be padded so that $\tilde{\rho}\in\bar{\mathbb{R}}^{MD}$. We have
\[
R_{\rho}^{L}\cap R_{\lambda}^{M}=R_{\tilde{\rho}}^{M}\cap R_{\lambda}^{M}=R_{\tilde{\rho}\wedge\lambda}^{M}.
\]
Hence, $\mathcal{R}$ is a $\pi$-system~\cite[Def. 1.1]{Klenke201308}.

Step 2) Let $\mathcal{M}$ denote the collection of sets $M\subseteq\mathbb{S}^{D}$
such that $\bar{p}_{\star}\left(\check{z}\in\partial M\right)=0$.
We have that 
\[
\bar{p}_{\star}\left(\check{z}\in\partial\mathbb{S}^{D}\right)=\bar{p}_{\star}\left(\check{z}\in\emptyset\right)=0.
\]
Also, if $M\subset N\in\mathcal{M}$, 
\begin{align*}
\bar{p}_{\star}\left(\check{z}\in\partial\left(N\setminus M\right)\right) & =\bar{p}_{\star}\left(\check{z}\in\partial N\cup\partial M\right)\\
 & \leq\bar{p}_{\star}\left(\check{z}\in\partial N\right)+\bar{p}_{\star}\left(\check{z}\in\partial M\right)=0,
\end{align*}
and, if $M_{k}\in\mathcal{M}$, $k\in\mathbb{N}$, are disjoint, and
$M=\bigcup_{k\in\mathbb{N}}M_{k}$, then, 
\[
\bar{p}_{\star}\left(\check{z}\in\partial\left(M\right)\right)\leq\sum_{k\in\mathbb{N}}\bar{p}_{\star}\left(\check{z}\in\partial M_{k}\right)=0.
\]
Hence, $\mathcal{M}$ is a $\lambda$-system~\cite[Def. 1.10]{Klenke201308}.

Step 3) Let $\delta\left(\mathcal{R}\right)$ be the smallest $\lambda$-system
containing $\mathcal{R}$. Since the distribution of $y$ is absolutely
continuous, then clearly so is its stationary distribution $\bar{p}\left(y\in S\right)$.
Hence, since the latter equals $\bar{p}_{\star}\left(z\in S\right)$,
we have $\mathcal{R}\subseteq\mathcal{M}$ and $\mathcal{M}$ is a
$\lambda$-system, then $\delta\left(\mathcal{R}\right)\subseteq\mathcal{M}$.
Thus, from Dynkin's $\pi$-$\lambda$ theorem~\cite[Th. 1.19]{Klenke201308},
$\delta\left(\mathcal{R}\right)=\sigma\left(\mathcal{R}\right)$.
But from~\cite[Th. 1.23]{Klenke201308} and the definition of $\mathcal{S}^{D}$,
$\sigma\left(\mathcal{R}\right)=\mathcal{S}^{D}$. So we have that
$\mathcal{S}^{D}\subseteq\mathcal{M}$ and the result follows. 
\end{IEEEproof}
\begin{lem}
\label{lem:AMSE} If $y$ is AMS, its distribution is absolutely continuous
and~(\ref{eq:undetect-cond}) holds, then $z$ is AMS and for all
$S\in\mathcal{S}^{D}$, 
\begin{equation}
\bar{p}\left(z\in S\right)=\bar{p}_{\star}\left(z\in S\right).\label{eq:conv-measures}
\end{equation}
\end{lem}
\begin{IEEEproof}
Let $t_{k}\in\mathbb{N}$, for $k=1,\cdots,K$, and $\tau=\left[t_{1},\cdots,t_{K}\right]^{\top}$.
Let also $\lambda_{k}\in\bar{\mathbb{R}}^{D}$, for $k=1,\cdots,K$,
and $\lambda=\left[\lambda_{1}^{\top},\cdots,\lambda_{K}^{\top}\right]^{\top}$.
Consider the set $D_{\lambda}^{\tau}\in\mathcal{S}^{D}$ defined by
\[
D_{\lambda}^{\tau}=\left\{ x\in\mathbb{S}^{D}:x_{t_{k}}\preceq\lambda\right\} .
\]
We have that $D_{\lambda}^{\tau}=R_{\rho}^{L}$, where $L=t_{k}$
and $\rho=\left[\rho_{1}^{\top},\cdots,\rho_{L}^{\top}\right]^{\top}$
with $\rho_{l}=\lambda_{l}$ if $l=t_{k}$ and $\infty\mathbf{1}$
otherwise. We then have that, for all $\tau$ and $\lambda$, 
\[
\lim_{T\rightarrow\infty}\bar{p}_{T}\left(z\in D_{\lambda}^{\tau}\right)=\lim_{T\rightarrow\infty}\frac{1}{T}\sum_{t=1}^{T}\mathcal{E}\left\{ \mathds{1}_{\left\{ \zeta_{t}^{(L)}\preceq\rho\right\} }\left(\rho\right)\right\} \overset{\text{(a)}}{=}\mathcal{E}\left\{ \lim_{T\rightarrow\infty}F_{T}^{(L)}\left(\rho\right)\right\} \overset{\text{(b)}}{=}F_{\star}^{(L)}(\rho)=\bar{p}_{\star}\left(z\in D_{\lambda}^{\tau}\right),
\]
where~(a) follows from Lebesgue's dominated convergence theorem and~(b)
follows from~(\ref{eq:undetect-cond}). It then follows from~\cite[Example 2.4]{Billingsley199907}
that the sequence of probability measures $\bar{p}_{T}\left(z\in\cdot\right):\mathcal{S}^{D}\rightarrow\mathbb{R}_{+}$
converges weakly to $\bar{p}_{\star}\left(z\in\cdot\right)$, i.e.,
for all $f\in C_{b}\left(\mathbb{S}^{D}\right)$ (the space of continuous
bounded functions on $\mathbb{S}^{D}$), 
\[
\lim_{T\rightarrow\infty}\int fd\bar{p}_{T}=\int fd\bar{p}_{\star}.
\]
It the follows from~\cite[Th. 2.1]{Billingsley199907}, that, for
each $S\in\mathcal{S}^{D}$, with $\bar{p}_{\star}\left(z\in\partial S\right)=0$,
\[
\lim_{T\rightarrow\infty}\bar{p}_{T}\left(z\in S\right)=\bar{p}_{\star}\left(z\in S\right).
\]
Then,~(\ref{eq:conv-measures}) follows from the above since, in
view of Lemma~\ref{lem:boundary}, $\bar{p}_{\star}\left(z\in\partial S\right)=0$
holds for all $S\in\mathcal{S}^{D}$.
\end{IEEEproof}
\begin{IEEEproof}[Proof of Theorem~\ref{thm:main}]
 \uline{Only if part}: If $z$ is asymptotically stealthy, since
$y$ is AMS, then clearly also is $z$. Also, since $\bar{p}_{\star}\left(z\in S\right)=\bar{p}\left(y\in S\right)$,
for all $S\in\mathcal{S}^{D}$, and $y$ is AMS and ergodic, it follows
from~\cite[Lemma 7.13]{Gray200908} that $z$ is ergodic. Then, from
the AMS ergodic theorem~\cite[Th. 8.1]{Gray200908}, 
\[
\lim_{T\rightarrow\infty}F_{T}^{(L)}(\rho)\overset{\text{w.p.1}}{=}\bar{\mathcal{E}}_{\star}\left\{ \mathds{1}_{R_{\rho}^{L}}\left(z\right)\right\} =F_{\star}^{(L)}(\rho),
\]
and~(\ref{eq:undetect-cond}) holds.

\uline{If part}: If~(\ref{eq:undetect-cond}) holds, since $y$
is AMS, the asymptotic stealthiness of $z$ follows immediately from
Lemma~\ref{lem:AMSE}.
\end{IEEEproof}

\subsection{The stationary linear Gaussian case\label{subsec:linear-Gaussian}}

In this subsection we specialize the result of Theorem~\ref{thm:main}
for the case in which~(\ref{eq:ss1})-(\ref{eq:ss2}) have the following
form
\begin{align*}
x_{t+1} & =Ax_{t}+w_{t},\\
y_{t} & =Cx_{t}+v_{t},
\end{align*}
with $w_{t}\sim\mathcal{N}\left(0,Q\right)$ and $v_{t}\sim\mathcal{N}\left(0,R\right)$.
We also assume that the system is in steady state, i.e., $x_{t}\sim\mathcal{N}\left(0,P\right)$,
with $P=APA^{\top}+Q$.

If we run a Kalman filter, in steady state we obtain 
\begin{align}
\hat{x}_{t+1|t} & =A\hat{x}_{t|t-1}+K\left(z_{t}-C\hat{x}_{t|t-1}\right),\label{eq:KF}\\
K & =A\Psi C^{\top}\left(C\Psi C^{\top}+R\right)^{-1},\nonumber 
\end{align}
where $\Psi$ is the solution of
\[
\Psi=A\Psi A^{\top}-A\Psi C^{\top}\left(C\Psi C^{\top}+R\right)^{-1}C\Psi A^{\top}+Q.
\]
Let $\hat{z}_{t|t-1}=C\hat{x}_{t|t-1}$, $\tilde{z}_{t}=z_{t}-\hat{z}_{t|t-1}$
and $\Gamma=C\Psi C^{\top}+R$. Let also 
\begin{equation}
\check{z}_{t}=\Gamma^{-1/2}\tilde{z}_{t}.\label{eq:norm-perror}
\end{equation}
Under nominal statistics, we have that $\Gamma$ is the covariance
of $\tilde{z}_{t}$. Thus, since the samples of $\tilde{z}$ are independent,
those of $\check{z}$ are independent and identically distributed
(i.i.d.), with $\check{z}_{t}\sim\mathcal{N}(0,\mathbf{I})$. In view
of this, it would be numerically more convenient if the condition
of Theorem~\ref{thm:main} could be given in terms of $\check{z}$.
This is done in the next corollary of Theorem~\ref{thm:main}. Let
$\check{F}_{T}^{(L)}$ be defined as in~(\ref{eq:CDFapprox}), but
with $\check{z}$ in place of $z$.
\begin{cor}
\label{cor:linear-gaussian} The process $z$ is asymptotically stealthy
if and only if, for all $L\in\mathbb{N}$ and $\rho\in\mathbb{R}^{LD}$,
\begin{equation}
\lim_{T\rightarrow\infty}\check{F}_{T}^{(L)}\left(\rho\right)\overset{\text{w.p.1}}{=}\Phi_{0,\mathbf{I}}(\rho).\label{eq:undetect-cond-normal}
\end{equation}
\end{cor}
\begin{IEEEproof}
We split the proof in steps:

Step 1) Clearly, if the statistics of $z$ are nominal, then so are
those of $\check{z}$ (i.e., it is i.i.d. with $\check{z}_{t}\sim\mathcal{N}(0,\mathbf{I})$).
The converse also holds since $z_{t}$ is fully determined by $\left(\check{z}_{s}:t\geq s\in\mathbb{Z}\right)$,
via the recursions
\[
\hat{x}_{t+1|t}=A\hat{x}_{t|t-1}+K\Gamma^{1/2}\check{z}_{t},\qquad z_{t}=C\hat{x}_{t|t-1}+\Gamma^{1/2}\check{z}_{t}.
\]
It is then straightforward to check that $\bar{p}\left(z\in S\right)=\bar{p}_{\star}\left(z\in S\right)$
if and only if $\bar{p}\left(\check{z}\in S\right)=\bar{p}_{\star}\left(\check{z}\in S\right)$.
Hence, asserting that $z$ has nominal stationary statistics is equivalent
to asserting that $\check{z}$ also has.

Step 2) Since $y$ is AMS, $z$ has nominal stationary statistics
if and only if it is asymptotically stealthy. The same conclusion
can be drawn for $\check{z}$. Then, combining these two facts with
the conclusion from Step~1), we obtain that $z$ is asymptotically
stealthy if and only if $\check{z}$ is so.

Step 3) Let $\check{y}$ be defined as $\check{z}$ but using $y$
in place of $z$. Clearly, $\check{y}$ satisfies the conditions of
$y$ in Theorem~\ref{thm:main}. It then follows from this theorem
that $\check{z}$ is asymptotically stealthy if and only if~(\ref{eq:undetect-cond-normal})
holds. The result then follows from the conclusion of Step~2).
\end{IEEEproof}
\begin{rem}
The family of tests~(\ref{eq:undetect-cond-normal}) can be understood
as a normality test. More precisely, as checking the following condition
\begin{equation}
\check{z}_{t}\text{ is }\mathcal{N}(0,\mathbf{I})\text{ and i.i.d}.\label{eq:cond-eq}
\end{equation}
Corollary~\ref{cor:linear-gaussian} asserts that this test enjoys
the property of being equivalent of the asymptotic stealthiness of
the process $z$. However, notice that not every normality test run
on the sequence $\check{z}_{t}$ may enjoy this property, and therefore
usable for assessing asymptotic stealthiness.
\end{rem}

\section{Attack examples whose detection requires Theorem~\ref{thm:main}\label{sec:relevant}}

Checking the condition of Corollary~\ref{cor:linear-gaussian} essentially
means checking that the stationary probability distribution $\bar{p}$
of $\check{z}$ equals the nominal one $\bar{p}_{\star}$, i.e., that
under the distribution $\bar{p}$,~(\ref{eq:cond-eq}) holds. As
it is known, from a theoretical point of view, checking that a block
of samples has joint standard normal distribution is a stronger requirement
than doing some other more practical checks, e.g., for pairwise independence
or uncorrelation. However, the question arise as to whether, for the
purposes of detecting an attack, it is really necessary to carry out
a full distribution check, or if instead, a simpler test would be
enough. In this section we provide two examples showing how an asymptotically
detectable attack can pass undetected if a test checking only for
uncorrelation or pairwise independence is used. This supports our
claim that checking the condition of Theorem~\ref{thm:main} is indeed
needed.

\subsection{Checking for uncorrelation\label{subsec:method-puncorr}}

By combining a normality test~\cite{thode2002testing}, together
with a test for uncorrelation~\cite[\S 14.2]{wasserman2004all},
we can verify the following condition
\begin{equation}
\check{z}_{t}\sim\mathcal{N}\left(0,\mathbf{I}\right)\text{ and \ensuremath{\bar{\ensuremath{\mathcal{E}}}}}_{\star}\left\{ \check{z}_{t}\check{z}_{s}^{\top}\right\} =0,\forall t\neq s\in\mathbb{N}.\label{eq:cond-white}
\end{equation}
However, condition~(\ref{eq:cond-eq}) is stronger that~(\ref{eq:cond-white}),
in the sense that the former requires that $\check{z}_{t}$ and $\check{z}_{s}$
are statistically independent, rather than uncorrelated, when $t\ne s$.
In this section we describe an attack example which can be detected
by a method verifying~(\ref{eq:cond-eq}), but not by one verifying~(\ref{eq:cond-white}).

Suppose that we feed the output $y$ to the Kalman filter~(\ref{eq:KF}).
Let $\check{y}_{t}\sim\mathcal{N}\left(0,\mathbf{I}\right)$ denote
the resulting normalized prediction error, obtained as in~(\ref{eq:norm-perror}).
Let $\tau\in\mathbb{N}$, $0<\upsilon<1$ and $\gamma_{t}$ be an
i.i.d. sequence of binary random variables with $p\left(\gamma_{t}=0\right)=p\left(\gamma_{t}=1\right)=0.5$.
Let $r_{0}\sim\mathcal{N}\left(0,\mathbf{I}\right)$ and 
\begin{eqnarray}
r_{t} & = & \upsilon r_{t-\tau}+\sqrt{1-\upsilon^{2}}\check{y}_{t},\qquad\check{z}_{t}=\gamma_{t}r_{t}.\label{attack_model1}
\end{eqnarray}
Since $\check{y}_{t}\sim\mathcal{N}\left(0,\mathbf{I}\right)$ is
i.i.d., it is straightforward to see that $r_{t}\sim\mathcal{N}\left(0,\mathbf{I}\right)$.
Hence $\check{z}_{t}\sim\mathcal{N}\left(0,\mathbf{I}\right)$. Also,
if $t\neq s$, 
\[
\mathcal{E}\left\{ \check{z}_{t}\check{z}_{s}^{\top}\right\} =\left(1-\upsilon^{2}\right)\mathcal{E}\left\{ \gamma_{t}\right\} \mathcal{E}\left\{ \gamma_{s}\right\} \mathcal{E}\left\{ r_{t}r_{s}\right\} =0.
\]
Hence, the process $\check{z}$ satisfies~(\ref{eq:cond-white}).
However, since 
\[
\check{z}_{t}=\gamma_{t}\left(\upsilon r_{t-\tau}+\sqrt{1-\upsilon^{2}}\check{y}_{t}\right),\quad\check{z}_{t-\tau}=\gamma_{t-\tau}r_{t-\tau},
\]
the vector $\left[\check{z}_{t}^{\top},\check{z}_{t-\tau}^{\top}\right]^{\top}$
is clearly not Gaussian. Hence, $\check{z}$ does not satisfy~(\ref{eq:cond-eq}).
Since the attacker knows $y_{s}$, for all $s\leq t$, it can always
build the attacking signal $z_{t}$ such that the normalized prediction
error $\check{z}$ at the receiver equals the one described above.
Such an attack can be detected by~(\ref{eq:cond-eq}) but not by~(\ref{eq:cond-white}).

\subsection{Checking for pairwise independence\label{subsec:method-pind}}

A combination of a normality test~\cite{thode2002testing} with a
test for pairwise independence~\cite[\S 15]{wasserman2004all},~\cite{bakirov2006multivariate}
permit checking the following condition
\begin{equation}
\left[\check{z}_{t}^{\top}\check{z}_{s}^{\top}\right]^{\top}\sim\mathcal{N}\left(0,\mathbf{I}\right),\forall t\neq s.\label{eq:cond-pair}
\end{equation}
As it is known, assessing that~(\ref{eq:cond-eq}) holds is not equivalent
to assessing~(\ref{eq:cond-pair}). This is because pairwise independence
does not imply joint independence in general. We describe below an
attacking scheme which would be detected by~(\ref{eq:cond-eq}),
but not by~(\ref{eq:cond-pair}).

Let the measurement dimension $D=1$. As before, we feed the output
$y$ to the Kalman filter~(\ref{eq:KF}), and let $\check{y}_{t}\sim\mathcal{N}\left(0,1\right)$
denote the normalized prediction error. Draw $\left[\check{z}_{0},\check{z}_{-1}\right]$
from the distribution $\mathcal{N}(0,\mathbf{I})$. Then, for $t\in\mathbb{N}$,
we compute 
\begin{equation}
\check{z}_{t}=\begin{cases}
\check{y}_{t}, & t\text{ even},\\
\mathrm{sign}\left(\check{z}_{t-1}\check{z}_{t-2}\right)\left|\check{y}_{t}\right|, & t\text{ odd}.
\end{cases}\label{eq:pair1}
\end{equation}

We first analyze pairwise independence. If $t$ is even $\check{z}_{t}$
is obviously independent of $\check{z}_{s}$, for all $s\neq t$.
So we assume that $t$ is odd. Suppose that at time $t$, the vector
$\left[\check{z}_{t-1},\check{z}_{t-2}\right]$ has distribution $\mathcal{N}\left(0,\mathbf{I}\right)$.
We have 
\[
p\left(\check{z}_{t},\check{z}_{t-1}\right)=p\left(\check{z}_{t}|\check{z}_{t-1}\right)p\left(\check{z}_{t-1}\right).
\]
Now 
\[
p\left(\check{z}_{t}=\beta|\check{z}_{t-1}\right)=\begin{cases}
\frac{1}{2}p\left(\left|\check{y}_{t}\right|=\beta\right), & \beta>0\\
\frac{1}{2}p\left(-\left|\check{y}_{t}\right|=\beta\right), & \beta\leq0
\end{cases}=\phi_{0,1}\left(\beta\right).
\]
Also, $\check{z}_{t-1}\sim\mathcal{N}(0,1)$. Hence $\left[\check{z}_{t},\check{z}_{t-1}\right]\sim\mathcal{N}\left(0,\mathbf{I}\right)$.
By symmetry, we also have that $\left[\check{z}_{t},\check{z}_{t-2}\right]\sim\mathcal{N}\left(0,\mathbf{I}\right)$.
Since clearly $\left[\check{z}_{t},\check{z}_{s}\right]\sim\mathcal{N}\left(0,\mathbf{I}\right)$
for any even $s$, it remains to be shown that $\left[\check{z}_{t},\check{z}_{2s+1}\right]\sim\mathcal{N}\left(0,\mathbf{I}\right)$,
for all $s$. This follows immediately from~(\ref{eq:pair1}), since
$\check{z}_{t-2}$ is independent of $\check{z}_{2s+1}$. Then, by
induction on $t$, we have that~(\ref{eq:pair1}) holds for all $t$
and $s$.

Now, clearly, if $t$ is even, $\left[\check{z}_{t},\check{z}_{t-1},\check{z}_{t-2}\right]\sim\mathcal{N}\left(0,\mathbf{I}\right)$,
However, for any odd $t$, $\check{z}_{t}\check{z}_{t-1}\check{z}_{t-2}\geq0$.
Then, 
\begin{equation}
\left[\check{z}_{t},\check{z}_{t-1},\check{z}_{t-2}\right]\nsim\mathcal{N}\left(0,\mathbf{I}\right).\label{eq:pair2}
\end{equation}
Hence, while $\check{z}$ is clearly AMS, in view of~(\ref{eq:pair2}),
it does not satisfy~(\ref{eq:cond-eq}). We can then draw the same
conclusions as those in Section~\ref{subsec:method-puncorr}.

\section{Attack detection algorithm\label{sec:algorithm}}

An attack detection algorithm cannot be readily obtained from Theorem~\ref{thm:main}
or Corollary~\ref{cor:linear-gaussian}. This is because this result
requires carrying out the uncountable family of tests~(\ref{eq:undetect-cond}).
Moreover, each of these tests requires an infinite number of attacked
samples. Hence, they cannot detect sporadic attacks that last a finite
time interval. Nevertheless, numerically tractable algorithms can
be readily derived using these results. In this section we derive
one such algorithm, for the case of linear systems with Gaussian noises,
based on the result from Corollary~\ref{cor:linear-gaussian}. More
precisely, we propose a test which, for each time step, produces a
single statistic which summarizes the outcome of a finite number of
tests~(\ref{eq:undetect-cond}). A similar algorithm for detecting
attacks in a general non-linear system can be derived from the conditions
of Theorem~\ref{thm:main}.

In order to cope with the fact that the family of tests~(\ref{eq:undetect-cond})
is uncountable, we first point out that it is enough to consider the
largest $L$ that the available computational resources allow. Then,
for this fixed value of $L$, we define a set of sample points on
$\mathbb{R}^{LD}$ at which we will test condition~(\ref{eq:undetect-cond}).
We use $\rho_{i}=\left[\rho_{i,0}^{\top},\cdots,\rho_{i,L-1}^{\top}\right]^{\top}$,
$i=1,\cdots,I$, to denote these points, with $\rho_{i,l}\in\mathbb{R}^{D}$,
$l=0,\cdots,L-1$. Also, in order to go around the limitation that
the tests~(\ref{eq:undetect-cond}) can not detect sporadic attacks,
while in Corollary~\ref{cor:linear-gaussian} $\check{F}_{T}^{(L)}$
is computed starting from time $1$, we use a moving horizon of $T$
time samples. Hence, for each sample time $t$ we define the block
of samples $\check{\zeta}_{t}^{(L)}=\left[\check{z}_{t}^{\top},\cdots,\check{z}_{t+L-1}^{\top}\right]^{\top}$
within this horizon, and build its empirical CDF
\begin{align}
\check{F}_{t,T}^{(L)}\left(\rho\right) & =\frac{1}{T}\sum_{s=t-T+1}^{t}\mathds{1}_{\left\{ \check{\zeta}_{s}^{(L)}\preceq\rho\right\} }(\rho).\label{eq:phi_t}
\end{align}
We then sample $\check{F}_{t,T}^{(L)}$ and $\Phi_{0,\mathbf{I}}$
at the points $\rho_{i}$, $i=1,\cdots,I$, forming the vectors $u_{t,T}\in\mathbb{R}^{I}$
and $u_{\star}\in\mathbb{R}^{I}$, respectively, defined by 
\begin{align}
\left[u_{t,T}\right]_{i} & =\check{F}_{t,T}^{(L)}\left(\rho_{i}\right),\quad\text{and}\quad\left[u_{\star}\right]_{i}=\Phi_{0,\mathbf{I}}\left(\rho_{i}\right).\label{eq:u_T-u_bar}
\end{align}
We next define the following weighted difference between the above
vectors 
\[
v_{t,T}=T\left(u_{t,T}-u_{\star}\right)^{\top}\Sigma^{-1}\left(u_{t,T}-u_{\star}\right),
\]
where $\Sigma=\sum_{\tau\in\mathbb{Z}}\Sigma(\tau)$, with 
\[
\Sigma(\tau)=\bar{\mathcal{E}}_{\star}\left\{ \left(\xi_{\tau}-u_{\star}\right)\left(\xi_{0}-u_{\star}\right)^{\top}\right\} ,
\]
and $\xi_{t}\in\mathbb{R}^{I}$ given by
\[
\left[\xi_{t}\right]_{i}=\mathds{1}_{\left\{ \check{\zeta}_{t}^{(L)}\preceq\rho_{i}\right\} }\left(\rho_{i}\right)
\]
We then have the following result.

In the proof of Proposition~\ref{prop:alg} requires Lemma~\ref{lem:CLT},
which is a generalization of the central limit theorem to the case
of finite independent vector processes.
\begin{defn}
A random process $x$ is $M$-independent if $x_{t}$ and $x_{s}$
are independent whenever $t-s>M$. It is called finite-independent
if it is $M$-independent for some $M\in\mathbb{N}$. 
\end{defn}
\begin{lem}
\label{lem:CLT} Let $x$ be a stationary finite-independent vector
random process with $\mathcal{E}\left\{ \left\Vert x_{0}\right\Vert ^{12}\right\} <\infty$.
Let $\bar{x}=\mathcal{E}\left\{ x_{0}\right\} $, $\Sigma(\tau)=\mathcal{E}\left\{ \left(x_{\tau}-\bar{x}\right)\left(x_{0}-\bar{x}\right)^{\top}\right\} $
and $\Sigma=\sum_{\tau\in\mathbb{Z}}\Sigma(\tau)$. Let also 
\begin{equation}
\hat{x}_{T}=\frac{1}{T}\sum_{t=1}^{T}x_{t}.\label{eq:u-bar}
\end{equation}
Then 
\[
\lim_{T\rightarrow\infty}\sqrt{T}\left(\hat{x}_{T}-\bar{x}\right)\overset{\mathcal{D}}{=}\mathcal{N}\left(0,\Sigma\right).
\]
\end{lem}
\begin{IEEEproof}
The scalar version of the result is a simplified version of~\cite[Th. 27.4]{billingsley2008probability}.
The vector version follows by applying Cramér-Wold theorem~\cite[Th. 29.4]{billingsley2008probability}
to the scalar version. 
\end{IEEEproof}
\begin{prop}
\label{prop:alg} Under nominal statistics, 
\[
\lim_{T\rightarrow\infty}v_{t,T}\overset{\mathcal{D}}{=}\chi^{2}\left(I\right),\text{ for all }t\in\mathbb{Z},
\]
where $\chi^{2}\left(I\right)$ is a chi-squared distribution with
$I$ degrees of freedom.
\end{prop}
\begin{IEEEproof}
We can write $u_{t,T}$ and $u_{\star}$ as
\[
u_{t,T}=\frac{1}{T}\sum_{s=t-T+1}^{t}\xi_{t},\quad\text{and}\quad u_{\star}=\bar{\mathcal{E}}_{\star}\left\{ \xi_{0}\right\} .
\]
Let $w_{t,T}=\sqrt{T}\Sigma^{-1/2}\left(u_{t,T}-u_{\star}\right)$.
It follows from~(\ref{eq:phi_t}) that, under nominal statistics,
the sequence $\left(\xi_{t}:t\in\mathbb{N}\right)$ is $(L-1)$-independent
(i.e., $\xi_{t}$ and $\xi_{s}$ are independent whenever $t-s>L-1$).
Hence, in view of Lemma~\ref{lem:CLT}
\begin{equation}
\lim_{T\rightarrow\infty}w_{t,T}\overset{\mathcal{D}}{=}\mathcal{N}\left(0,\mathbf{I}\right).\label{eq:conv-normal}
\end{equation}
Now $v_{t,T}$ can be written as
\begin{equation}
v_{t,T}=\left\Vert w_{t,T}\right\Vert _{2}^{2}=\sum_{i=1}^{I}\left[w_{t,T}\right]_{i}^{2}.\label{eq:v_T}
\end{equation}
Then, the result follows from~(\ref{eq:conv-normal}) and the continuous
mapping theorem~\cite[Corollary 1 of Th. 25.7]{billingsley2008probability}.
\end{IEEEproof}
Let $H_{I}$ denote the CDF of a chi-squared distribution with $I$
degrees of freedom. We define the confidence $\psi_{t,T}$ of rejecting
the null hypothesis (i.e., of asserting that there is an attack) at
time $t$ by
\begin{equation}
\psi_{t,T}=H_{I}\left(v_{t,T}\right).\label{eq:stats}
\end{equation}
In view of Proposition~\ref{prop:alg}, $\psi_{t,T}$ is uniformly
distributed on $[0,1]$. Hence, for a given alarm triggering threshold
$0\leq\alpha\leq1$ on $\psi_{t,T}$, the false alarm rate $\pi_{t,T}$
at time $t$ (i.e., the probability of triggering an alarm when there
is no attack), is given by 
\[
\pi_{t,T}=p\left(\psi_{t,T}\geq\alpha\right)=1-\alpha.
\]

In order to compute $\psi_{t,T}$ we need expressions for $\xi_{t}$,
$u_{\star}$ and $\Sigma$. These are given in the next proposition.
\begin{prop}
For each $i,j=1,\cdots,I$, we have 
\begin{align}
\left[\xi_{t}\right]_{i} & =\prod_{l=0}^{L-1}\mathds{1}_{\left\{ \check{z}_{t+l}\preceq\rho_{i,l}\right\} },\;\text{and}\;\left[u_{\star}\right]_{i}=\prod_{l=0}^{L-1}\Phi_{0,\mathbf{I}}\left(\rho_{i,l}\right).\label{eq:alg}
\end{align}
Also, 
\begin{equation}
\Sigma=\sum_{t=-L+1}^{L-1}\Sigma(t),\label{eq:Sigma-sum}
\end{equation}
where
\begin{equation}
\Sigma(t)=\tilde{\Sigma}(t)-\bar{\Sigma},\label{eq:Sigma}
\end{equation}
with
\begin{equation}
\left[\bar{\Sigma}\right]_{i,j}=\prod_{\tau=0}^{L-1}\Phi_{0,\mathbf{I}}\left(\rho_{i,\tau}\right)\Phi_{0,\mathbf{I}}\left(\rho_{j,\tau}\right),\label{eq:Sigma-bar}
\end{equation}
and
\begin{equation}
\left[\tilde{\Sigma}(t)\right]_{i,j}=\begin{cases}
\prod_{\tau=0}^{-t-1}\Phi_{0,\mathbf{I}}\left(\rho_{i,\tau}\right)\Phi_{0,\mathbf{I}}\left(\rho_{j,L+t+\tau}\right)\prod_{\tau=0}^{L+t-1}\Phi_{0,\mathbf{I}}\left(\rho_{i,\tau-t}\wedge\rho_{j,\tau}\right), & -L+1\leq t<0,\\
\prod_{\tau=0}^{L-1}\Phi_{0,\mathbf{I}}\left(\rho_{i,\tau}\wedge\rho_{j,\tau}\right), & t=0,\\
\prod_{\tau=0}^{t-1}\Phi_{0,\mathbf{I}}\left(\rho_{i,L-t+\tau}\right)\Phi_{0,\mathbf{I}}\left(\rho_{j,\tau}\right)\prod_{\tau=0}^{L-t-1}\Phi_{0,\mathbf{I}}\left(\rho_{i,\tau}\wedge\rho_{j,t+\tau}\right), & 0<t\leq L-1.
\end{cases}\label{eq:Sigma-tilde}
\end{equation}
\end{prop}
\begin{IEEEproof}
Equation~(\ref{eq:alg}) follows straightforwardly from~(\ref{eq:u_T-u_bar}).
To show~(\ref{eq:Sigma-tilde}), recall that $\rho_{i}=\left[\rho_{i,0}^{\top},\rho_{i,1}^{\top},\ldots,\rho_{i,L-1}^{\top}\right]^{\top}$,
$i=1,\cdots,I$, are vectors on $\mathbb{R}^{LD}$. For all $i=1,\cdots,I$,
\[
\left[\xi_{t}\right]_{i}=\prod_{l=0}^{L-1}\mathds{1}_{\left\{ \check{z}_{t+l}\preceq\rho_{i,l}\right\} },\quad\text{and}\quad\left[u_{\star}\right]_{i}=\prod_{l=0}^{L-1}\Phi_{0,\mathbf{I}}\left(\rho_{i,l}\right).
\]
Then, for all $i,j=1,\cdots,I$, and $t\in\mathbb{Z}$, 
\begin{equation}
\left[\Sigma(t)\right]_{i,j}=\left[\tilde{\Sigma}(t)\right]_{i,j}-\left[\bar{\Sigma}\right]_{i,j},\label{eq:tmp}
\end{equation}
with
\begin{align*}
\left[\tilde{\Sigma}(t)\right]_{i,j} & =\bar{\mathcal{E}}_{\star}\left\{ \prod_{l=0}^{L-1}\mathds{1}_{\left\{ \check{z}_{t+l}\preceq\rho_{i,l}\right\} }\mathds{1}_{\left\{ \check{z}_{l}\preceq\rho_{j,l}\right\} }\right\} ,\\
\left[\bar{\Sigma}\right]_{i,j} & =\prod_{l=0}^{L-1}\Phi_{0,\mathbf{I}}\left(\rho_{i,l}\right)\Phi_{0,\mathbf{I}}\left(\rho_{j,l}\right).
\end{align*}

Suppose that $t\geq0$. We have
\begin{align}
\left[\tilde{\Sigma}(t)\right]_{i,j} & =\bar{\mathcal{E}}_{\star}\left\{ \prod_{\tau=0}^{t-1}\mathds{1}_{\left\{ \check{z}_{\tau}\preceq\rho_{j,\tau}\right\} }\prod_{\tau=0}^{L-t-1}\mathds{1}_{\left\{ \check{z}_{t+\tau}\preceq\rho_{i,\tau}\right\} }\mathds{1}_{\left\{ \check{z}_{t+\tau}\preceq\rho_{j,t+\tau}\right\} }\prod_{\tau=0}^{t-1}\mathds{1}_{\left\{ \check{z}_{L+\tau}\preceq\rho_{i,L-t+\tau}\right\} }\right\} \nonumber \\
 & =\prod_{\tau=0}^{t-1}\Phi_{0,\mathbf{I}}\left(\rho_{i,L-t+\tau}\right)\Phi_{0,\mathbf{I}}\left(\rho_{j,\tau}\right)\prod_{\tau=0}^{L-t-1}\Phi_{0,\mathbf{I}}\left(\rho_{i,\tau}\wedge\rho_{j,t+\tau}\right).\label{eq:aux-1}
\end{align}
and
\[
\left[\bar{\Sigma}\right]_{i,j}=\prod_{\tau=0}^{t-1}\Phi_{0,\mathbf{I}}\left(\rho_{i,L-t+\tau}\right)\Phi_{0,\mathbf{I}}\left(\rho_{j,\tau}\right)\prod_{\tau=0}^{L-t-1}\Phi_{0,\mathbf{I}}\left(\rho_{i,\tau}\right)\Phi_{0,\mathbf{I}}\left(\rho_{j,t+\tau}\right).
\]
Then, subtracting and factor out the common term we get
\[
\left[\Sigma(t)\right]_{i,j}=\left[\prod_{\tau=0}^{t-1}\Phi_{0,\mathbf{I}}\left(\rho_{i,L-t+\tau}\right)\Phi_{0,\mathbf{I}}\left(\rho_{j,\tau}\right)\right]\left[\prod_{\tau=0}^{L-t-1}\Phi_{0,\mathbf{I}}\left(\rho_{i,\tau}\wedge\rho_{j,t+\tau}\right)-\prod_{\tau=0}^{L-t-1}\Phi_{0,\mathbf{I}}\left(\rho_{i,\tau}\right)\Phi_{0,\mathbf{I}}\left(\rho_{j,t+\tau}\right)\right].
\]
The above equation shows that $\left[\Sigma(t)\right]_{i,j}=0$ for
all $t\geq L$.

In the above equations we considered the case $t\geq0$. This means
that the samples $\check{z}_{t},\cdots,\check{z}_{t+L-1}$ associated
with index $i$ are right-shifted with respect to those $\check{z}_{0},\cdots,\check{z}_{L-1}$
associated with index $j$. If we now consider the case $t\leq0$
(both cases coincide when $t=0$), the samples associated with $j$
will be right-shifted with respect to those of $i$. Since the process
$\check{z}$ is stationary, the equations will be the same as those
above but swapping the indexes $i$ and $j$ and taking $-t$ instead
of $t$. Hence we get 
\begin{equation}
\left[\tilde{\Sigma}(t)\right]_{i,j}=\prod_{\tau=0}^{-t-1}\Phi_{0,\mathbf{I}}\left(\rho_{i,\tau}\right)\Phi_{0,\mathbf{I}}\left(\rho_{j,L+t+\tau}\right)\prod_{\tau=0}^{L+t-1}\Phi_{0,\mathbf{I}}\left(\rho_{i,\tau-t}\wedge\rho_{j,\tau}\right),\label{eq:aux-2}
\end{equation}
and
\[
\left[\Sigma(t)\right]_{i,j}=\left[\prod_{\tau=0}^{-t-1}\Phi_{0,\mathbf{I}}\left(\rho_{i,\tau}\right)\Phi_{0,\mathbf{I}}\left(\rho_{j,L+t+\tau}\right)\right]\left[\prod_{\tau=0}^{L+t-1}\Phi_{0,\mathbf{I}}\left(\rho_{i,\tau-t}\wedge\rho_{j,\tau}\right)-\prod_{\tau=0}^{L+t-1}\Phi_{0,\mathbf{I}}\left(\rho_{i,\tau-t}\right)\Phi_{0,\mathbf{I}}\left(\rho_{j,\tau}\right)\right],
\]
which shows that $\left[\Sigma(t)\right]_{i,j}=0$ for all $t\leq-L$.
Hence~(\ref{eq:Sigma-sum}) follows. Equations~(\ref{eq:Sigma})
and~(\ref{eq:Sigma-bar}) follows from~(\ref{eq:tmp}). And~(\ref{eq:Sigma-tilde})
follows from~(\ref{eq:aux-1}) and~(\ref{eq:aux-2}).
\end{IEEEproof}
The proposed method assesses the presence of an attack by measuring
the squared distance between the nominal and empirical CDFs of $\check{\zeta}_{1}^{(L)}$.
Since the domain of these functions is $\mathbb{R}^{LD}$, their distance
is measured over the grid sample points $\rho_{i}\in\mathbb{R}^{LD}$,
$i=1,\cdots,I$. In order to complete the description of the method,
we need a criterion for choosing these points. To this end, we apply
the generalized Lloyd's algorithm~\cite[S 11.3]{GershoGray199111}
to the nominal probability distribution of $\check{\zeta}_{1}^{(L)}$,
i.e., $\mathcal{N}\left(0,\mathbf{I}\right)$. We then obtain the
algorithm summarized in Algorithm~\ref{alg:1}.

\begin{algorithm}[H]
\textbf{Initialization:} choose $T,L,I\in\mathbb{N}$ and a threshold
$\alpha>0$. 
\begin{enumerate}
\item Run Lloyd's algorithm on the $LD$-dimensional distribution $\mathcal{N}\left(0,\mathbf{I}\right)$,
to obtain $\rho_{i}\in\mathbb{R}^{LD}$, $i=1,\cdots,I$. 
\item Using the points $\rho_{i}$, $i=1,\cdots,I$, compute $u_{\star}$
and $\Sigma$. 
\end{enumerate}
\textbf{Main loop:} at time $t$, let $\tau=t-L+1$ and run the following
steps.
\begin{enumerate}
\item Run the Kalman filter~(\ref{eq:KF}) to obtain $\hat{z}_{t|t-1}$. 
\item Compute $\check{z}_{t}$ using~(\ref{eq:norm-perror}). 
\item Compute $\check{\zeta}_{\tau}^{(L)}$ using $\check{z}_{\tau},\cdots,\check{z}_{t}$. 
\item Compute $u_{t,T}$ using $\check{\zeta}_{\tau-T+1}^{(L)},\cdots,\check{\zeta}_{\tau}^{(L)}$. 
\item Using $u_{t,T}$ compute $v_{t,T}$. 
\item Compute $\psi_{t,T}$ using $v_{t,T}$ and~(\ref{eq:stats}). 
\item Trigger an alarm if $\psi_{t,T}\geq\alpha$.
\end{enumerate}
\caption{Attack detection test}

\label{alg:1} 
\end{algorithm}

\begin{rem}
Algorithm~\ref{alg:1} is an approximation of the family of tests~(\ref{eq:undetect-cond-normal}).
The key property of this algorithm is that this approximation can
be made arbitrarily accurate by increasing the values of $T$, $L$
and $I$.
\end{rem}
\begin{rem}
\label{rem:time-horizon}Concerning the choice of the time horizon
$T$, while a large value increases the accuracy and therefore the
sensitivity of the detector to the presence of an attack, it also
decreases the detection speed, i.e., the time that the detector takes
to react to the occurrence of an attack. Hence, in a practical implementation,
$T$ need to be chosen to accommodate a trade off between sensitivity
and speed. It is also possible to achieve a combination of high detection
speed to strong attacks and high sensitivity for detecting subtle
attacks with slow detection speed, by running in parallel two instances
of the same method, one with a small time horizon and the other with
a large one.
\end{rem}

\section{Simulation\label{sec:simulate}}

In this section we illustrate our proposed method. Since this method
checks that the \emph{joint} statistics (JS) of a block of contiguous
samples equal their nominal values, we refer to it as JS. We compare
the JS method with other two. The first one is the method described
in Section~\ref{subsec:method-pind}, which checks for the normality
as well as pairwise independence (NPI) of samples of $\check{z}$.
To do so, the method compares the joint CDFs of the vector $\left[\check{z}_{t}^{\top},\check{z}_{t-l}^{\top}\right]^{\top}$,
for all values $l=1,\cdots,L-1$, using a procedure similar to the
one described in Section~\ref{sec:algorithm}. This yields the $L-1$
statistics $v_{t,T}^{(l)}$, $l=1,\cdots,L-1$, which are computed
as in~(\ref{eq:v_T}). We refer to this method as NPI. The second
method is the one described in~\cite[eqs. (6)-(7)]{mo2016performance}.
This method checks that the second order (SO) statistics of samples
of the prediction error $\tilde{z}$ equal their nominal values. In
our notation, it defines 
\[
v_{t,T}=\tilde{z}_{t}^{\top}\Gamma\tilde{z}_{t}\sim\chi^{2}\left(D\right).
\]
We refer to this method as SO.

To do the comparison, we use a system with $A=0.98$, $C=1$, $R=0.1$
and $Q=0.1.$ Also, for the JS and NPI methods we use $I=100$, $L=3$
and $T=100$.

\begin{figure}[h]
\begin{centering}
\includegraphics[width=0.49\columnwidth]{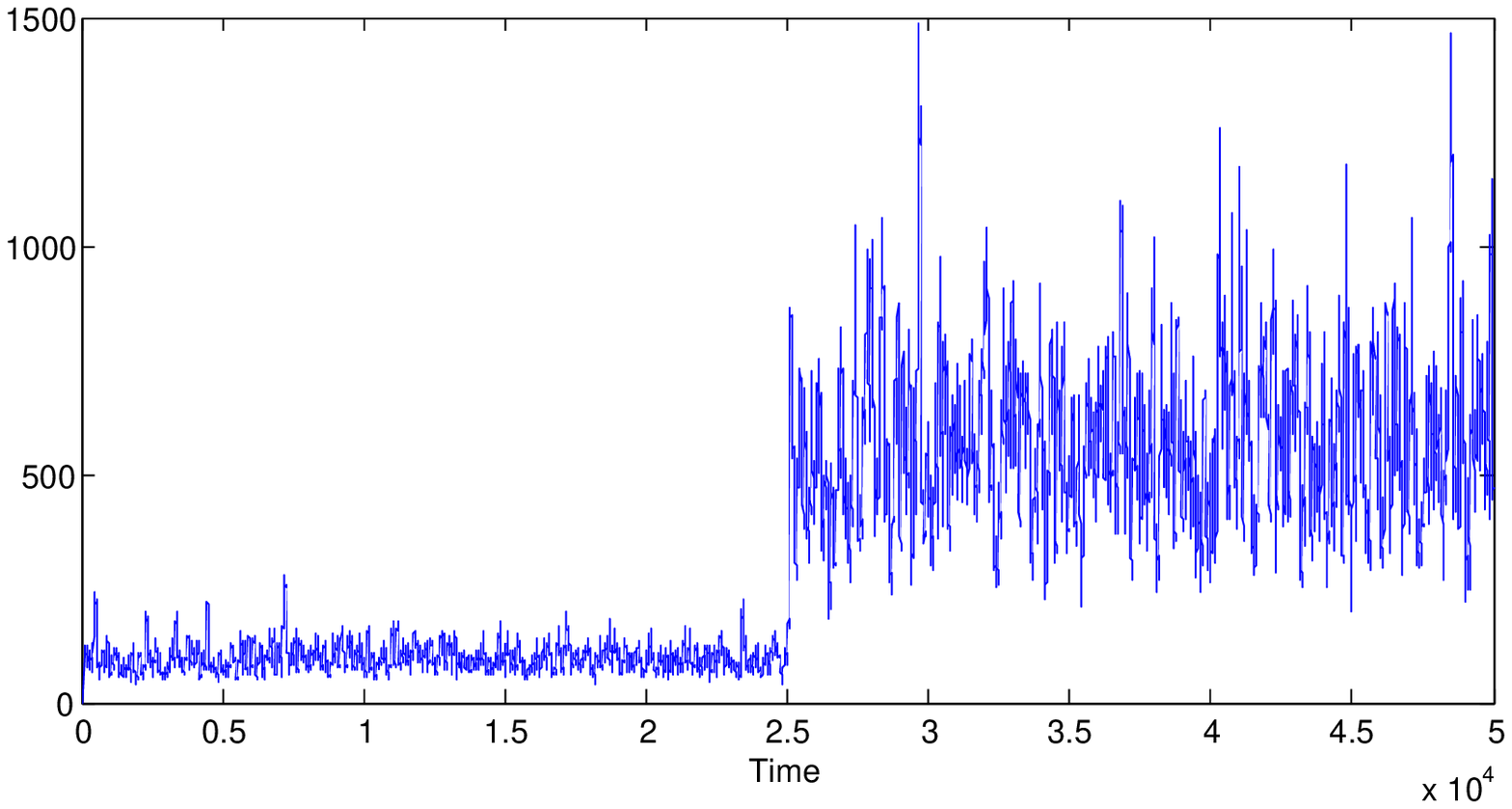} \includegraphics[width=0.49\columnwidth]{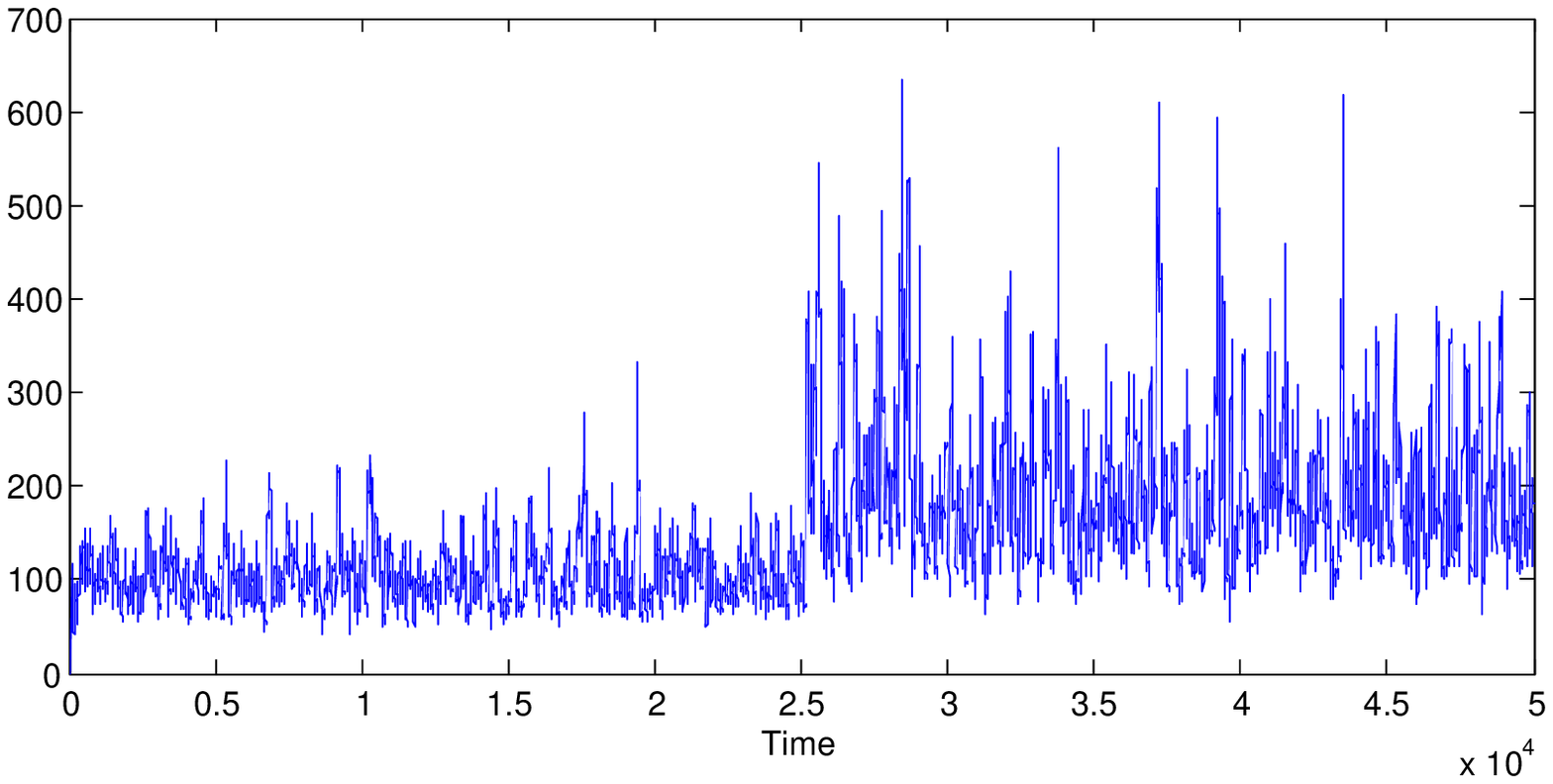}
\par\end{centering}
\caption{Evolution of $v_{t,T}$ yield by the JS detection method under attack~(\ref{attack_model1})
(left) and attack~(\ref{eq:pair1}) (right).}
\label{attack1_vT} 
\end{figure}

\begin{figure}[h]
\begin{centering}
\includegraphics[width=0.49\columnwidth]{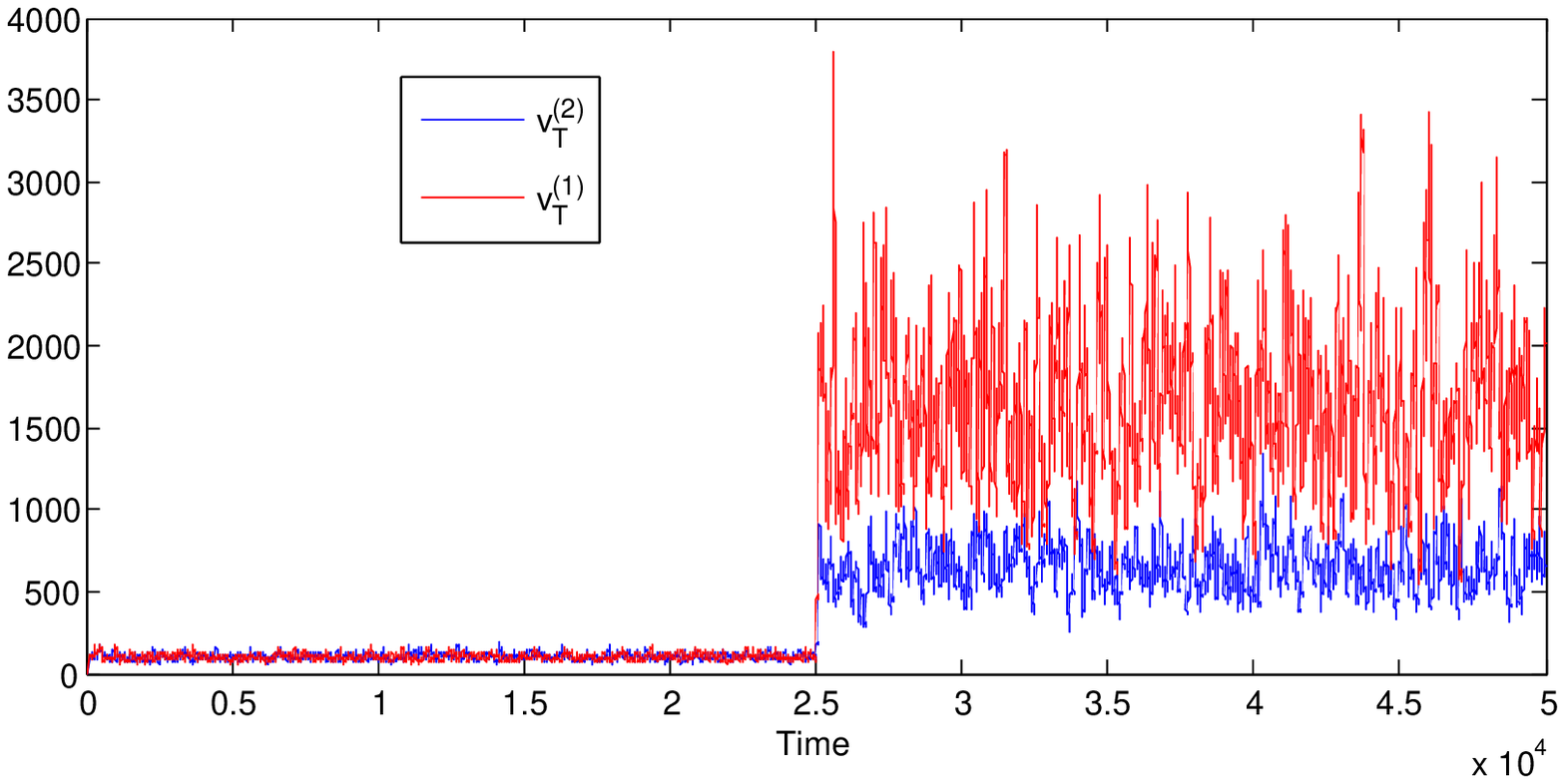} \includegraphics[width=0.49\columnwidth]{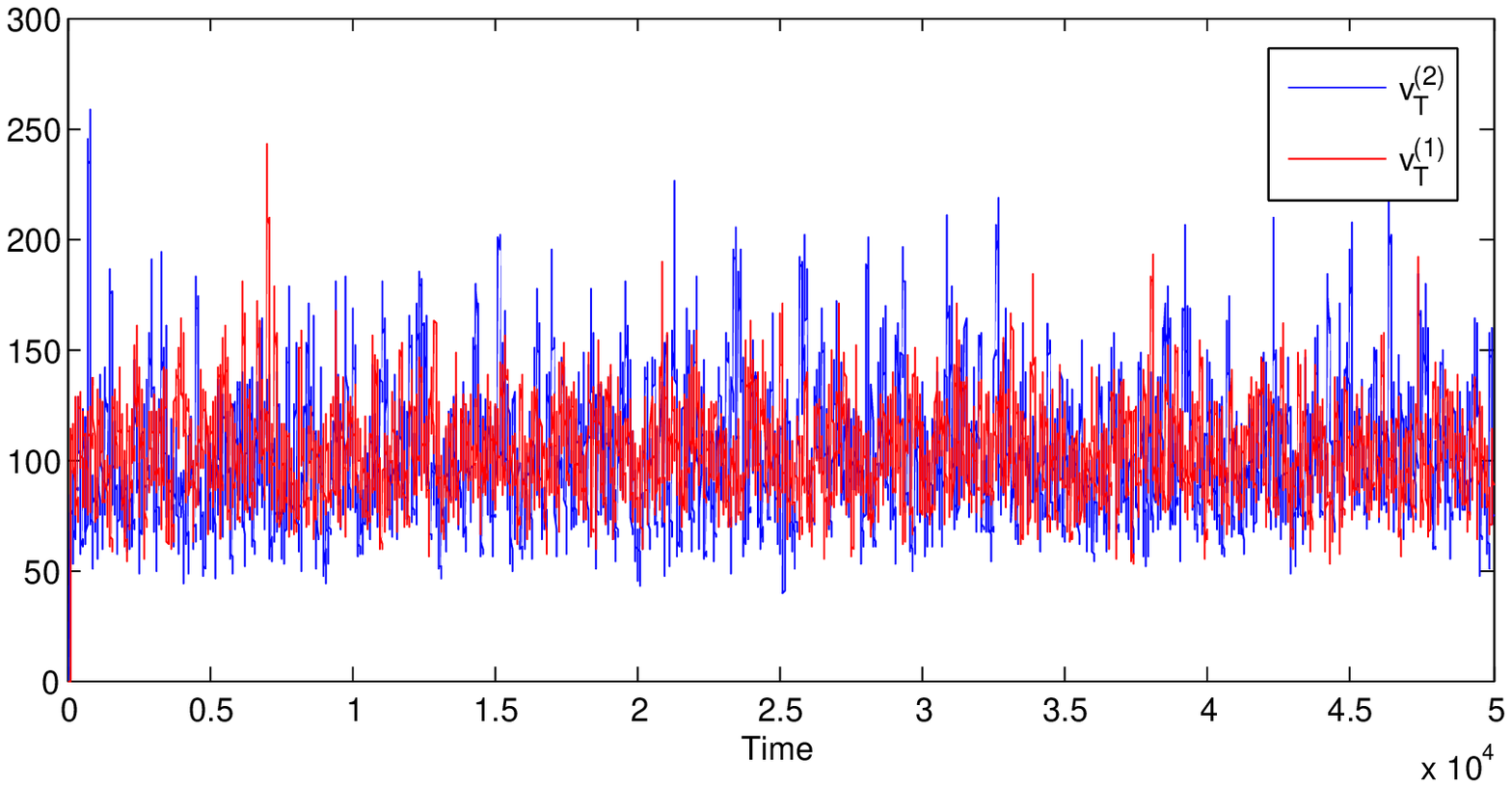}
\par\end{centering}
\caption{Evolution of $v_{t,T}^{(1)}$ and $v_{t,T}^{(2)}$ yield by the NPI
detection method under attack~(\ref{attack_model1}) (left) and attack~(\ref{eq:pair1})
(right).}
\label{attack1_vT_pariwise} 
\end{figure}

\begin{figure}[h]
\begin{centering}
\includegraphics[width=0.49\columnwidth]{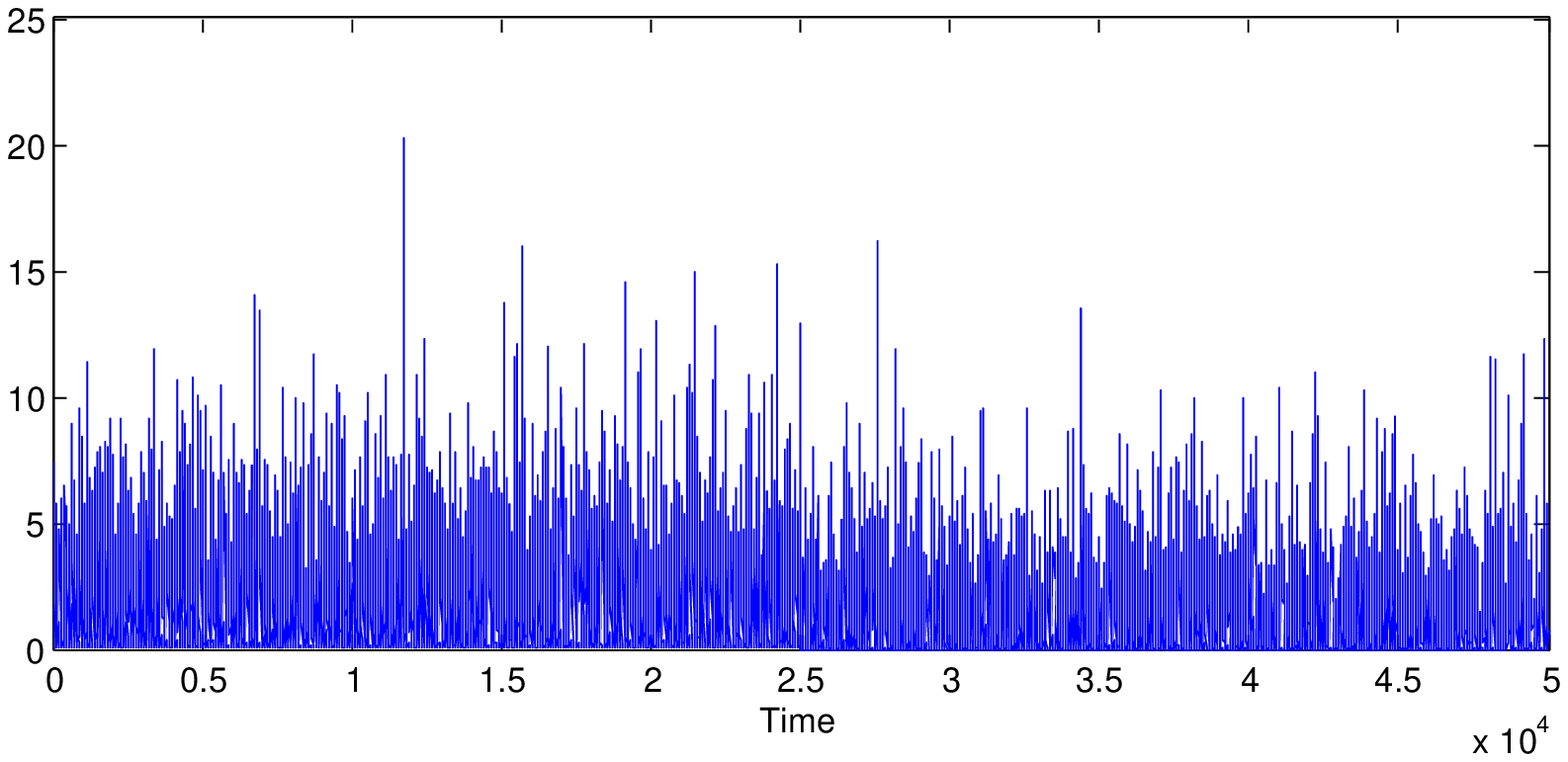} \includegraphics[width=0.49\columnwidth]{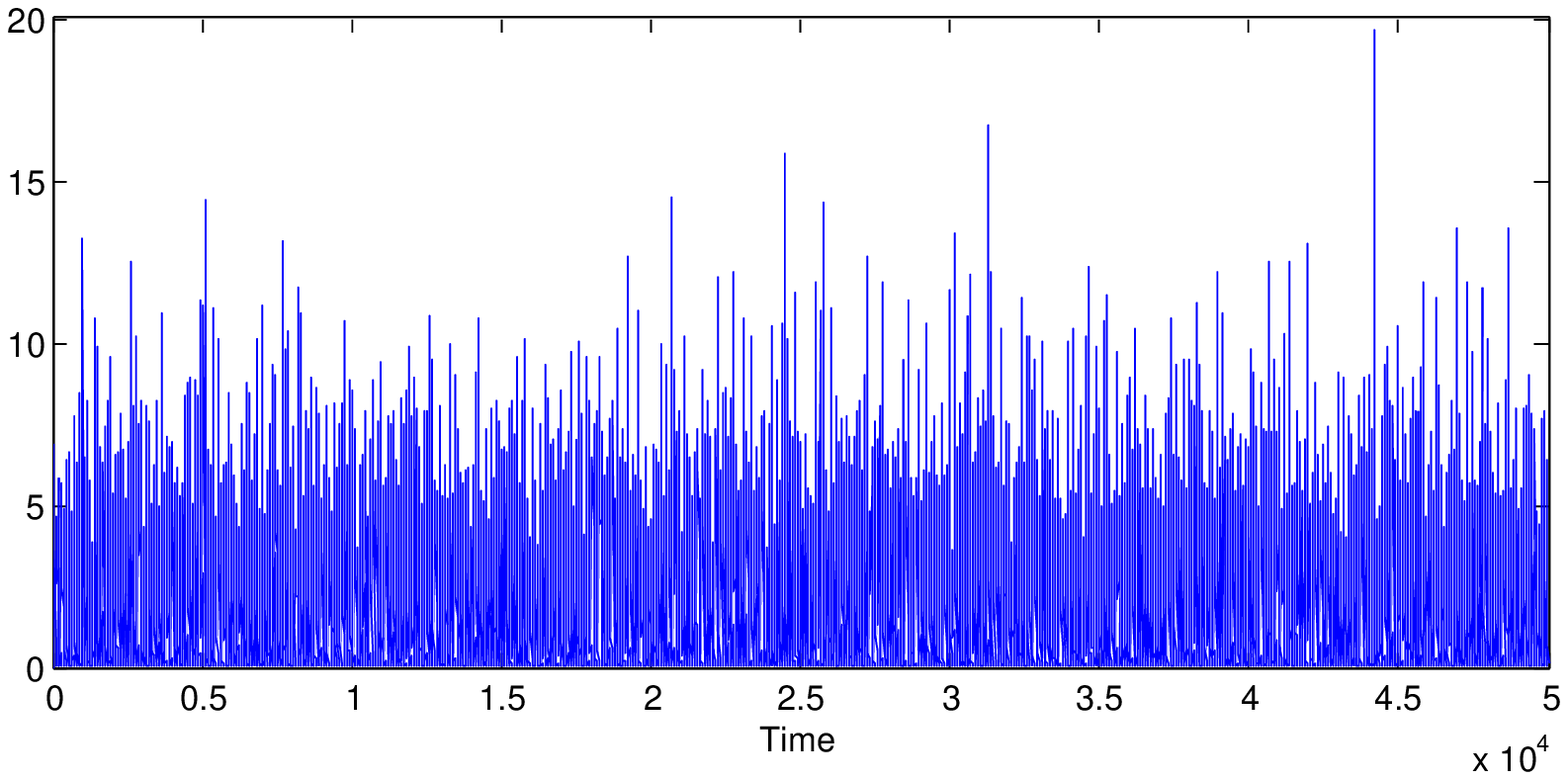}
\par\end{centering}
\caption{Evolution of $v_{t,T}$ yield by the SO detection method~\cite{mo2016performance}
under attack~(\ref{attack_model1}) (left) and attack~(\ref{eq:pair1})
(right).}
\label{attack1_gk} 
\end{figure}

In the first experiment we consider the attack described in~(\ref{attack_model1}),
with $\tau=1$ and $\alpha=1/\sqrt{2}$. As described in Section~\ref{subsec:method-puncorr},
this attack introduces statistical dependence between samples of $\check{z}$
which are $\tau$ samples away from each other. However, these samples
remain uncorrelated. The left sides of Figures~\ref{attack1_vT},~\ref{attack1_vT_pariwise}
and~\ref{attack1_gk} show the values of the statistic $v_{t,T}$
for the methods JS, NPI and SO, respectively. We see how the SO method
is unable to detect the appearance of the attack at time $t=25\times10^{3}$.

In the second experiment, we consider the attack described in~(\ref{eq:pair1}).
As explained in that section, this attack introduces statistical dependence
between three consecutive samples of $\check{z}$, while leaving all
samples from $\check{z}$ being pairwise independent. The values of
$v_{t,T}$ for the methods JS, NPI and SO, are shown at the right
sides of Figures~\ref{attack1_vT},~\ref{attack1_vT_pariwise} and~\ref{attack1_gk},
respectively. We see that in this case, only the JS method is able
to detect the appearance of the attack.
\begin{rem}
As stated in Corollary~\ref{cor:linear-gaussian}, checking~(\ref{eq:undetect-cond-normal}),
for arbitrarily large $L$ and all $\rho$, guarantees the detection
of all asymptotically detectable attacks. However, recall that algorithm
JS only checks~(\ref{eq:undetect-cond-normal}) for a fixed $L$
and a finite number of points $\rho_{i}\in\mathbb{R}^{LD}$, $i=1,\cdots,I$.
Hence, certain asymptotically detectable attacks can escape the detection
of algorithm JS. While the set of escaping attacks can be made arbitrarily
small by increasing $L$ and $I$, there exists always the possibility
that some attack can escape algorithm JS but not other algorithm like
SO.
\end{rem}

\section{Conclusion\label{sec:conclusion}}

We studied the attack detection problem on stochastic cyber-physical
systems. We introduced the definition of asymptotic detectable attacks,
as the set of attacks that can be detected by some method based on
the knowledge of a single realization, and with probability bigger
than zero over the space of realizations. We also characterized this
set by providing a necessary and sufficient condition for stochastic
detectability. Using this condition, we derived a practically realizable
attack detection algorithm. We present simulation results showing
how our algorithm can detect attacks that cannot be detected by some
simpler methods.

\bibliographystyle{unsrt}
\bibliography{refs}

\end{document}